\documentclass{appolb}
\usepackage{graphicx}
\usepackage{amsmath}

\newcommand{\D}{\mathrm{d}}
\newcommand{\df}[1]{\delta\left(#1\right)}
\newcommand{\eps}{\epsilon}
\newcommand{\nn}{\nonumber}
\newcommand{\G}[1]{\operatorname{H}_{#1}}
\newcommand{\order}{\mathcal{O}}
\renewcommand{\sp}[2]{\,#1\!\cdot\!#2\,}
\newcommand{\spr}[2]{#1\!\cdot\!#2}
\newcommand{\x}{\;}
\newcommand{\zb}{\zeta_2}
\newcommand{\zc}{\zeta_3}
\newcommand{\zd}{\zeta_4}

\begin{document}
\title{
\begin{flushright}
{\normalsize DESY 17-191}
\end{flushright}
\vspace{5mm}
Fuchsia and Master Integrals for Energy-Energy Correlations at NLO in QCD%
}
\author{Oleksandr~Gituliar\thanks{Presented at Matter to the Deepest Conference in Podlesice (Poland) \mbox{3--8}~Sep~2017 and at RADCOR Conference in St.~Gilgen (Austria) 24--29~Sep~2019.} and Sven-Olaf~Moch, 
\address{II. Institut f\"ur Theoretische Physik, Universit\"at Hamburg,\\Luruper Chaussee 149, D-22761 Hamburg, Germany}
}
\maketitle

\begin{abstract}
In this talk we discuss some aspects of the analytical calculation of energy
correlations in electron-positron annihilation at a next-to-leading order in QCD.  
Our primary focus is on the most difficult task: the calculation of master
integrals for real-emission contributions, which are functions of
two dimensionless variables and the dimensional regulator. 
We use a method of differential equations and their so-called epsilon-form 
which is constructed with the help of the {\sc Fuchsia} program based on Lee's algorithm. 
\end{abstract}


\section{Introduction}
The energy-energy correlation (EEC) is an observable proposed back in 1978 in \cite{BBEL78}
to test the consistency of quantum chromodynamics (QCD), 
when QCD was still considered as 
{\it "an appealing candidate for the field theory of hadronic interactions"}.
In those times, EEC was considered a very convenient observable to measure
from the experimental point of view, since no definition of a jet axis is needed. 
From the theoretical point of view, EEC is an observable free of mass
singularities, as can be easily deduced from its definition,
\begin{equation}
\label{eq:eec}
  \mathrm{\Sigma}(\xi) = \sum_{a,b}\x \int \mathrm{dPS}\x \frac{E_a E_b}{Q^2}\x \sigma\big(e^+ + e^- \to a+b+X\big)\x \delta\left(\xi - \cos\theta_{ab}\right)
  \, ,
\end{equation}
since the phase space integration of the cross section $\sigma$ with the
measure $\mathrm{dPS}$ weighted with the energies $E_a$ and $E_b$ of partons
$a$ and $b$ eliminates all divergences arising from soft and collinear
particles production.
Nevertheless, the fact that individual diagrams may still diverge requires the
use of some regularization scheme, which we choose to be dimensional
regularization in $m=4-2\epsilon$ space-time dimensions.

Over a time span of almost 40 years EEC was thoroughly studied.
Among the recent results is the so far most accurate fixed-order calculation
at next-to-next-to-leading order (NNLO)~\cite{DDK16} in QCD. 
This result was subsequently combined with resummation of large logarithms to
next-to-next-to-leading logarithmic (NNLL) accuracy and low-energy
hadronization effects~\cite{TKS17}, which allowed for a detailed comparison
with available data from LEP.

The focus of our study is twofold. 
First, we aim to obtain a fully analytical result at next-to-leading order
(NLO) in QCD which to our surprise is not yet available for the moment. 
This is in contrast to $N=4$ super-Yang-Mills theory, where analytical 
expressions for the NLO corrections have recently been derived~\cite{BHK13} 
and one expects that the latter results correspond to those in QCD 
(after the usual identification of color factors) 
as far as the polylogarithms of highest weight are concerned.
Second, we wish to extend the application of Lee's algorithm~\cite{Lee15} and
the {\sc Fuchsia} program~\cite{GM16,GM17} beyond univariate problems.
To our knowledge, there are no examples of such usage in the literature so far 
while the method is not limited to univariate problems only.
With the NLO integrals for real-emission contributions in EEC at hand, 
we have an example of a multivariate system of differential equations~\cite{Kot91a,Kot91b}, 
demonstrate its solution with {\sc Fuchsia} 
to obtain the so-called epsilon form~\cite{Henn13}, 
and discuss a method to fix the necessary integration constants.

\section{The Method}
\begin{figure}[htb]
\centerline{%
  \includegraphics[width=0.175\paperwidth]{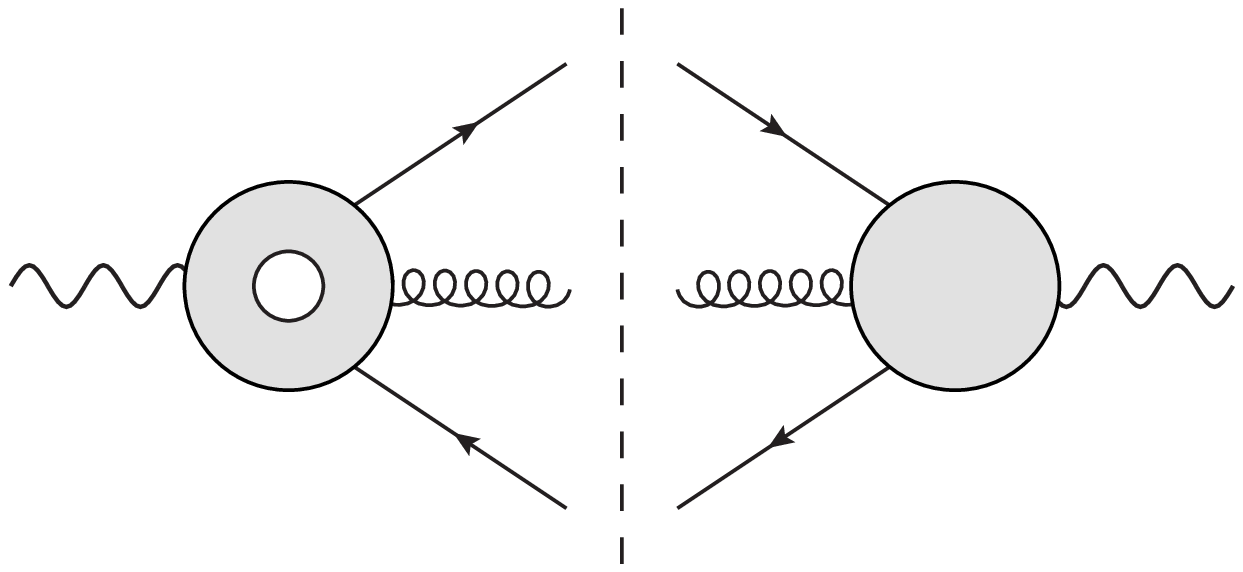}
  \includegraphics[width=0.175\paperwidth]{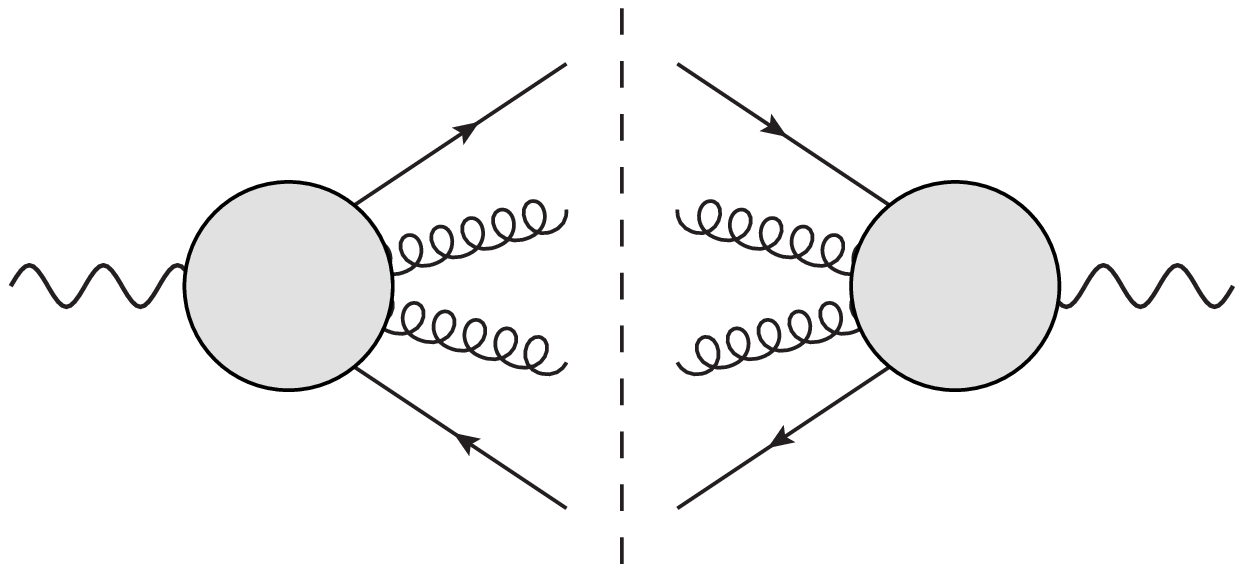}
  \includegraphics[width=0.175\paperwidth]{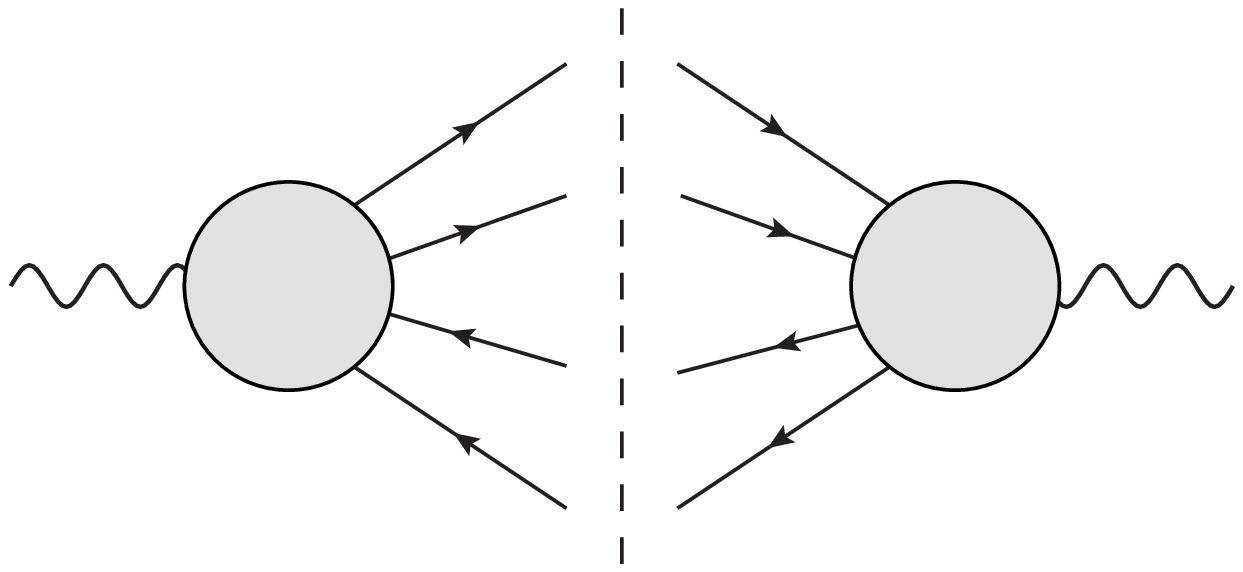}}
\caption{The virtual and the two distinct real emission sub-processes contributing to EEC at NLO.}
\label{fig:eec}
\end{figure}

The contributions to EEC at NLO in QCD arise from the three electron-positron
annihilation sub-processes depicted in Fig.~\ref{fig:eec}. 
The first ones, the virtual contribution, are rather simple to calculate by
directly integrating corresponding matrix elements available
from~\cite{GGG01}.
The last two, on the other hand, form the most difficult part of our calculation.
For clarity, we consider here only the first one with gluons,
corresponding to the center one Fig.~\ref{fig:eec}:  
\begin{equation}
  e^+ + e^- \to \gamma^*(q)\to { q(p_1)} + \bar{q}(p_2) + { g(p_3)} + g(p_4), \qquad p_i^2=0
  \, .
\end{equation}
Here, $p_i$ denote the light-like final-state particle momenta, 
while $q$ (with $q^2 = Q^2 > 0$) is the sum of the initial $e^+$ and $e^-$ momenta.
In addition, we restrict ourselves among all possible combinations of
correlated particles $ab$, i.e., $q\bar{q}$, $qg$, $\bar{q}g$, and $gg$,  
to the $qg$-correlation only.
The corresponding squared matrix elements are generated with {\sc FeynArts}~\cite{Hahn00} and {\sc FormCalc}~\cite{HP98} based on {\sc FORM}~\cite{Ver00}.
All the remaining correlations can be calculated in a similar way.

Let us start with an explicit parametrization of the phase-space in eq.~\eqref{eq:eec} 
in terms of kinematic invariants, which are suitable for a further integral
reduction by means of integration-by-parts (IBP)~\cite{CT81}. 
To that end, we set $q^2 = 1$ and introduce a new, more convenient, angular variable
\begin{equation}
  z = \frac{1-\xi}{2} = \frac{\spr{p_1}{p_3}}{2\x\spr{q}{p_1}\x\spr{q}{p_3}}
  \, ,
\end{equation}
which, however, still contains non-linear kinematic invariants not yet
suitable for an IBP reduction.
To fix this, we introduce a new parameter
\begin{equation}
x = 2\spr{q}{p_1}
\end{equation}
which leads to the final parametrization of the semi-inclusive three-particle phase-space
\begin{equation}
  \label{eq:ps3xz}
  \D \mathrm{PS}(3;x,z) =
    x\sp{q}{p_3}\x
    \df{1-x-(q-p_1)^2}
    \df{x\x z\sp{q}{p_3} - \sp{p_1}{p_3}}
    \D \mathrm{PS}(3)
\, ,
\end{equation}
where the corresponding inclusive three-particle phase-space 
$\D \mathrm{PS}(3)$ is given by
\begin{equation}
  \label{eq:ps3}
\D \mathrm{PS}(3) =
    \D^m p_1 \, \delta(p_1^2) \;
    \D^m p_2 \, \delta(p_2^2) \;
    \D^m p^3 \, \delta(p_3^2) \;
    \delta{\left((q-p_1-p_2-p_3)^2\right)}
    \, .
\end{equation}
With this parametrization IBP rules for the corresponding real-emission phase
space integrals can be applied to derive differential equations for the set of
master integrals.
These master integrals now depend on the two scale-less parameters $x$ and $z$, which is an added complication. 
However, we will see later this can be easily resolved.

\subsection{IBP Reduction for $F(x,z,\eps)$}
In the IBP reduction of real-emission phase space integrals 
all delta functions in eq.~\eqref{eq:ps3} 
are treated as cut propagators according to the Cutkosky rules.
We use the program {\sc LiteRed}~\cite{Lee12,Lee13} for this task, which
produces a family of 11 master integrals $F_i(x,z,\eps)$ defined as 
\begin{equation}
  \label{eq:MI}
\begin{aligned}
  F_1 & = \{\}
  &
  F_2 & = \{2\}
  &
  F_3 & = \{2,2\}
  &
  F_4 & = \{2,6\}
  \\
  F_5 & = \{1,2\}
  &
  F_6 & = \{5\}
  &
  F_7 & = \{1,4,5\}
  &
  F_8 & = \{2,3,4\}
  \\
  F_9 & = \{2,5\}
  &
  F_{10} & = \{3,5\}
  &
  F_{11} & = \{2,4,5\}
\end{aligned}
\, ,
\end{equation}
where phase space integration according to eq.~\eqref{eq:ps3xz} is understood.
The integrals $F_2$, $F_3$, and $F_4$ form a coupled sub-system and the denominators are defined~\footnote{Note, that $D_6$ corresponds to the on-shell delta function $\delta(p_1^2)$ and denotes its additional power when used in the parametrization of $F_4$, e.g, $\delta^2(p_1^2)$.} as
\begin{equation}
\begin{aligned}
  D_1 & = (p_2+p_3)^2 
  &
  D_2 & = (q-p_2)^2
  &
  D_3 & = (q-p_1-p_2)^2
  \\
  D_4 & = (q-p_1-p_3)^2 
  &
  D_5 & = (q-p_2-p_3)^2
  &
  D_6 & = p_1^2
  \quad D_7 = \sp{q}{p_3}
\end{aligned}
\, .
\end{equation}

\subsection{Differential Equations for $F(x,z,\eps)$}
With the complete set of IBP reduction rules at hand we can easily construct a
system of differential equations. 
In our approach we consider ordinary differential equations, where only one
variable is free and the rest are treated as symbolic constants.
At this stage we choose $z$ as a free variable, which leads to a system of equations
\begin{equation}
  \frac{\D F(x,z,\eps)}{\D z} = A(x,z,\eps)\x F(x,z,\eps)
\end{equation}
with singular points (alphabet) in
\begin{equation}
  0,\quad 1,\quad \frac{1}{x},\quad \frac{1}{x\x(x-2)}
  \, .
\end{equation}
To solve this system we find its epsilon form \cite{Henn13} 
with {\sc Fuchsia}~\cite{GM16,GM17} (see also \cite{Pra17,Mey16,Mey17}) and then integrate it to any order in
$\eps$ using a recursive definition of hyperlogarithms (see~\cite{Pan15} and
references therein). 
The resulting new system takes the form
\begin{equation}
   \frac{\D \hat{F}(x,z,\eps)}{\D z} = \eps\x B(x,z)\x \hat{F}(x,z,\eps)
   \, ,
\end{equation}
with the relation between new and old bases given as
\begin{equation}
  F(x,z,\eps) = T(x,z,\eps)\x\hat{F}(x,z,\eps)
  \, ,
\end{equation}
where the matrix $T(x,z,\eps)$ has been found automatically by {\sc Fuchsia}.

For clarity, let us provide here the solution to one of master integrals in 
eq.~\eqref{eq:MI}, 
\begin{equation}
  \label{eq:f4}
  \begin{aligned}
  & F_4(x,z,\eps) =
    \frac{1}{15\x\eps^2}\bigg({ C_3^0(x)} - 2 { C_2^0(x)}\bigg)
  + \frac{1}{30\x\eps}
    \Bigg(
        \bigg( 15 { C_1^0(x)} + 4{ C_2^0(x)}
    \\ & \quad - 6x{ C_3^0(x)} - 2x{ C_4^0(x)} \bigg) \G{0}(z) + \bigg(\frac{15}{1-x}{ C_1^0(x)} + 2 { C_3^0(x)}
    \\ & \quad - 2x { C_4^0(x)} \bigg) \G{1}(z)
      + \bigg(\frac{15(x-2)}{1-x} { C_1^0(x)} + 20 { C_2^0(x)} + 4x { C_4^0(x)}
    \\ & \quad
       + 2(3x-7) { C_3^0(x)}\bigg) \G{1/x}(z) + \frac{2\big(13xz^2-17xz+3x+z\big)}{xz(1-z)} { C_3^0(x)}
    \\ & \quad
      - \frac{15(1-2z)}{xz(1-z)} { C_1^0(x)} + \frac{4(13xz-1)}{xz} { C_2^0(x)} + \frac{2(xz-2z+1)}{z(1-z)} { C_4^0(x)}
    \\ & \quad
      - 4 { C_2^1(x)} + 2 { C_3^1(x)} \Bigg) + \order\big(\eps^0\big)
      \, ,
  \end{aligned}
\end{equation}
where the polylogarithms $\G{\vec{w}}$ are recursively defined as
\begin{equation}
  \label{eq:polylog}
  \G{a,\vec{w}}(z) = \int_0^z \frac{\D z'}{z'-a}\x \G{\vec{w}}(z')
  \, .
\end{equation}

\subsection{Integration Constants for $F(x,z,\eps)$}
The unknown integration constants $C(x)$ in eq.\eqref{eq:f4} are functions of $x$ and still need to be found.
For this purpose we employ the following relation
\begin{equation}
  \label{eq:fx}
  \boxed{F_i^\star(x,\eps) = \int_0^1 \D z \x f_i(z)\x F_i(x,z,\eps).}
\end{equation}
Here, the idea is the following: 
If we know both sides of this relation we can derive a system of linear
equations with $C(x)$ as unknown functions 
by requiring that coefficients in front of identical polylogarithms 
on both sides of eq.~\eqref{eq:fx} are equal.
Solutions of this system provide those unknown integration constants $C(x)$.

The integration on the right-hand side of eq.~\eqref{eq:fx} can be done with
the help of the {\sc HyperInt} package~\cite{Pan14}.
However, particular attention should be paid at this stage 
since some of these integrals diverge (like in the case of $F_4(x,z,\eps)$ above).
We account for this by introducing additional $z$-dependent pre-factors $f(z)$
for each of the master integrals in eq.~\eqref{eq:MI}. 
These are listed in the table:
$$
\begin{tabular}{|c | c | c | c | c | c | c | c | c | c | c | c |}
  \hline
  $i$ & 1 & 2 & 3 & 4& 5 & 6 & 7 & 8 & 9 & 10 &11 \\
  \hline
  $f_i$ & 1 & 1 & $z$ & $z\x(1-z)$ & 1 & $z^2$ & $1-z$ & $z$ & $z$ & $z$ & $z\x(1-z)$ \\
  \hline
\end{tabular}
$$
In this way, all the integrals on the right-hand side of eq.~\eqref{eq:fx}
become finite and can be integrated.

\subsection{IBP Reduction for $G(x,\eps)$}
For the left-hand side of eq.~\eqref{eq:fx} we can introduce a new IBP basis,
since the integration over $z$ eliminates one delta function in
eq.~\eqref{eq:ps3xz} and leads to a new three-particle phase-space given by 
\begin{equation}
\label{eq:ps3new}
  \D \mathrm{PS}(3;\x x)
=
\int_0^1 \D z \x \D \mathrm{PS}(3;\x x,z)
=
  \D \mathrm{PS}(3) \x \df{1-x-(q-p_1)^2}
  \, .
\end{equation}

We find the IBP reduction rules for this basis, again using {\sc LiteRed}, which leads to 12 master integrals defined as
\begin{equation}
\label{eq:gs}
\begin{aligned}
  G_1 & = \{\}
  &
  G_2 & = \{2\}
  &
  G_3 & = \{7\}
  &
  G_4 & = \{2,7\}
  \\
  G_5 & = \{2,\not 6,7\}
  &
  G_6 & = \{1,2\}
  &
  G_7 & = \{2,3,4,7\}
  &
  G_8 & = \{5,7\}
  \\
  G_9 & = \{2,4,5\}
  &
  G_{10} & = \{2,4,5,7\}
  &
  G_{11} & = \{3,5,7\}
  &
  G_{12} & = \{1,4,5,7\}
\end{aligned}
\, .
\end{equation}
where phase space integration according to eq.~\eqref{eq:ps3new} is again implied.
In this way we can express the left-hand side of eq.~\eqref{eq:fx} by using only these masters.
For example,
\begin{equation}
\label{eq:f4x}
\begin{aligned}
  & F_4^\star(x,\eps) =
      \frac{(2-3\eps)\x\big(x+5\eps x-2\eps^2\x(8-7x)\big)}{4\x\eps^2\x x^2\x(4-5x)} \x G_1(x,\eps)
    \\ & \quad
    + \frac{x\x(1-x)+\eps\x\big(16-33x+15x^2\big)-\eps^2\x\big(48-82x+26x^2\big)}{4\x\eps\x x^2\x(4-5x)} \x G_2(x,\eps)
    \\ & \quad
    + \frac{(1-2\eps)\x\big(x-2\eps\x(2-x)\big)}{4\x\eps\x x\x(4-5x)} \x G_3(x,\eps)
    \\ & \quad
    - \frac{4-7x+2x^2+\eps\x x\x(4-2x)}{4\x x\x(4-5x)} \x G_4(x,\eps)
    - \frac{3\x x\x(1-x)}{4\x(4-5x)} \x G_5(x,\eps)
    \, .
\end{aligned}
\end{equation}

\subsection{Differential Equations for $G(x,\eps)$}
Next, we compute the master integrals in eq.~\eqref{eq:gs} 
by solving the corresponding system of differential equations, but this time in $x$ variable.
As before, we find its epsilon form using {\sc Fuchsia} and solve the
resulting system to desired order in $\eps$ by applying the recursive definition
of polylogarithms in eq.~\eqref{eq:polylog}.
For example, the solution to $G_5$ looks as follows:
\begin{equation}
\label{eq:g5}
\begin{aligned}
  & G_5(x,\eps) =
    \frac{2}{x\x\eps^2}\Bigg(30 { C^0_1} + 6 { C_2^0} + 6 { C_3^0} + (14 - 35 x) { C_4^0} - 2 { C_5^0}\Bigg)
 + \frac{1}{x\x\eps} \Bigg( + 60 { C_1^1}
\\ &
 -390 { C_1^0} - 78 { C_2^0}
 + 12 { C_2^1} - 78 { C_3^0}
 + 12 { C_3^1} - (182 - 455 x) { C_4^0}
 + (28 - 70 x) { C_4^1}
\\ &
 + 26 { C_5^0} - 4 { C_5^1}
 + \Big((60 - 120 x) { C_1^0} + (132 - 144 x) { C_2^0}
 - (48 - 36 x) { C_3^0}
\\ &
 - (112 - 84 x) { C_4^0}
 + (16 - 12 x) { C_5^0}\Big) \G{0}(x)
 + \Big((-480 + 120 x) { C_1^0}
 - (96 - 144 x) { C_2^0}
\\ &
 - 36(1 + x) { C_3^0}
 + (-224 + 336 x) { C_4^0} + (-8 + 12 x) { C_5^0}\Big) \G{1}(x)\Bigg)
 + \order\big(\eps^0\big)
\, .
\end{aligned}
\end{equation}
In this case, the integration constants $C_i^{0,1}$ are now just numbers.

\subsection{Integration Constants for $G(x,\eps)$}
In order to obtain the integration constants for the integrals $G(x,\eps)$ we use the same technique as for $F(x,z,\eps)$ before.
We define a relation
\begin{equation}
  \label{eq:gx}
  \boxed{G_i^\star(\eps) = \int_0^1 \D x \x g_i(x)\x G_i(x,\eps)}
\end{equation}
and integrate its right-hand side with {\sc HyperInt} using now $x$-dependent 
pre-factors $f(x)$ to regularize the denominators in $x$. 
For each of the master integrals in eq.~\eqref{eq:gs} they are given in the table:
$$
\begin{tabular}{|c | c | c | c | c | c | c | c | c | c | c | c | c |}
  \hline
  $i$ & 1 & 2 & 3 & 4& 5 & 6 & 7 & 8 & 9 & 10 & 11 & 12 \\
  \hline
  $g_i$ & 1 & 1 & 1 & 1 & $x$ & 1 & $(1-x)^2$ & 1 & $x$ & $x(1-x)$ & $1-x$ & $1-x$ \\
  \hline
\end{tabular}
\x.
$$

\subsection{IBP Reduction for $H(\eps)$}
For the last iteration, that is the computation of the left-hand side of eq.~\eqref{eq:gx} 
we introduce a new IBP basis, since, as before, the integration eliminates one
delta function and the new phase-space becomes
\begin{equation}
\int_0^1 \D x \x \D \mathrm{PS}(3;\x x)
=
  \D \mathrm{PS}(3)
  \, ,
\end{equation}
cf. eq.~\eqref{eq:ps3}.
We find the necessary IBP reduction rules for this basis again with {\sc LiteRed}
and only two master integrals remain,
\begin{equation}
\label{eq:h}
\begin{aligned}
  H_1 & = \{\}
  &
  H_2 & = \{1,2,7\}
  \, .
\end{aligned}
\end{equation}
These are sufficient to determine the left-hand side of eq.~\eqref{eq:gx}.
For example, we find
\begin{equation}
\label{eq:g5x}
\begin{aligned}
  G_5^\star(\eps) & = - \frac{2\x(2-3\eps)\x(3-4\eps)\x\big(1-7\eps+30\eps^2-36\eps^3\big)}{3\x\eps^2\x\big(1-5\eps+6\eps^2\big)} \x H_1(\eps)
  \, .
\end{aligned}
\end{equation}

\section{Results}
We can now summarize all the results for the master integrals $H(\eps)$, $G(x,\eps)$ and $F(x,z,\eps)$ 
in this order by subsequently substituting the solutions for $H(\eps)$ back into $G(x,\eps)$ and 
$G(x,\eps)$ back into $F(x,z,\eps)$. 

\subsection{Master Integrals $H(\eps)$}
The integrals in eq.~\eqref{eq:h} coincide with fully inclusive phases-space integrals.
The first one is known in the literature~\cite{GGH03}, i.e.,
\begin{multline}
  \label{eq:h1}
  H_1(\eps)
  =
      \frac{1}{12}
    + \frac{59}{72} \eps
    + \left(\frac{2\x263}{432} - \frac{2}{3}\x\zb \right) \eps^2
    + \left(\frac{72\x023}{2\x592} - \frac{59}{9}\x\zb - \frac{13}{6}\x\zc\right) \eps^3
  \\
    + \left(\frac{2\x073\x631}{15\x552} - \frac{2\x263}{54}\x\zb - \frac{767}{36}\x\zc + \frac{1}{12}\x\zd\right) \eps^4
    + \order\big(\eps^5\big)
  \, ,
\end{multline}
and we calculate the second one explicitly as:
\begin{equation}
  \label{eq:h2}
  H_2(\eps)
  = -\frac{4\zc}{\eps} - 42\zd + \order(\eps)
  \, .
\end{equation}

\subsection{Master Integrals $G(x,\eps)$}
Next, we substitute eqs.~\eqref{eq:h1}--\eqref{eq:h2} into eq.~\eqref{eq:g5x} which delivers us the left-hand side of eq.~\eqref{eq:gx}.
The right-hand side was integrated with {\sc HyperInt} and we can determine
the integration constants in eq.~\eqref{eq:g5}. 
The result is given by
$$\begin{aligned}
  & G_{5}(x,\eps) =
\frac{1}{3x} \Bigg[
  - \frac{1}{\eps^2}
  + \frac{\G0(x)+4\G1(x)}{\eps}
  - (7-6x)\x \G{0,0}(x) - 2(5-3x)\x \G{0,1}(x)
\\ \nn & \quad
  - 2(2+3x)\x \G{1,0}(x) - 2(5+3x)\x \G{1,1}(x) - 2(1-3x)\x\zb + \Bigg( \big(61-54x\big) \G{0,0,0}(x)
\\ \nn & \quad
  + (46-36x) \G{0,0,1}(x) + 4\G{0,1,0}(x) + 28 \G{0,1,1}(x) - 18 x \G{0,1,1}(x) + 4 \G{1,0,0}(x)
\\ \nn & \quad
  + 18 x \G{1,0,0}(x) + 16 \G{1,0,1}(x)+4\G{1,1,0}(x) + 36 x \G{1,1,0}(x) + 10 \G{1,1,1}(x)
\\ \nn & \quad
  + 54 x \G{1,1,1}(x)+ \zb\x\big(38\G{0}(x)-36x\G{0}(x)-16\G{1}(x)\big) + \big(36-18x\big)\x\zc\Bigg)\x\eps \Bigg] + \order\big(\eps^2\big)
  \, .
\end{aligned}$$
In this way we can obtain the entire set of integrals in eq.~\eqref{eq:gs}.

\subsection{Master Integrals $F(x,z,\eps)$}
Finally, we can compute the left-hand side of eq.~\eqref{eq:fx} by using IBP
reduction rules similar to eq.~\eqref{eq:f4x} and find the remaining unknown integration
constants $C(x,\eps)$ in eq.~\eqref{eq:f4}. 
As an example, the result for $F_{11}$ reads
\begin{equation}
\begin{aligned}
  & F_{11}(x,z,\eps) =
\frac{4}{3 (1-x) x^2 (1-z) z} \Bigg[
   \frac{3 + x - 4 x z}{2\eps^2}
  + \frac{1}{\eps} \Bigg(-2 (3 + x - 4 x z) \G1(x)
\\ \nn & \quad
 - (6 - x - 5 x z) \G0(x) - (3 + x - 4 x z) \G0(z) - (3 - x - 2 x z) \G1(z)
\\ \nn & \quad
 + 6 (1 - x z) \G{1/x}(z) \Bigg)
  + 9 (1-x) \G{0, 1}(x) + 8 (3 + x - 4 x z) \G{1,1}(x) + (24
\\ \nn & \quad
 - 7 x - 17 x z) \G{0,0}(x) + \Big(2 (6 - x - 5 x z) \G0(z) + 4 (3 - x - 2 x z) \G{1}(z)
\\ \nn & \quad
 - 15 (1 - x z) \G{\frac{1}{x}}(z) - 3 (1 - x z) \G{\frac{1}{x(2-x)}}(z)\Big) \G0(x) + \G{0}(1-x) \Big(5 (3 + x
\\ \nn & \quad
 - 4 x z) \G{0}(x) + 4 (3 + x - 4 x z) \G{0}(z) + 4 (3 - x - 2 x z) \G{1}(z) - 21 (1
\\ \nn & \quad
 - x z) \G{\frac{1}{x}}(z) - 3 (1 - x z) \G{\frac{1}{x(2-x)}}(z)\Big) + 2 (3 + x - 4 x z) \G{0,0}(z) + 2 (3 - x 
\\ \nn & \quad
 - 2 x z) \G{0,1}(z) - 12 (1 - x z) \G{0, \frac{1}{x}}(z) + 2 (3 - x - 2 x z) \G{1,0}(z) + 2 (3 - 2 x
\\ \nn & \quad
 - x z) \G{1,1}(z)
  - 6 (2 - x - x z) \G{1, \frac{1}{x}}(z) - 9 (1 - x z) \G{\frac{1}{x}, 0}(z) - 6 (1 - x z)
\\ \nn & \quad
 \times \G{\frac{1}{x}, 1}(z) + 15 (1 - x z) \G{\frac{1}{x}, \frac{1}{x}}(z)
  - 3 (1 - x z) \G{\frac{1}{x(2-x)}, 0}(z) + 3 (1 - x z)
\\ \nn & \quad
 \times \G{\frac{1}{x(2-x)},\frac{1}{x}}(z) + 2 (3 - 5 x + 2 x z) \zb \Bigg] + \order(\eps)
 \, .
\end{aligned}
\end{equation}
This is the final result computed to sufficient depth in $\eps$ 
which can then be further used to calculate the EEC defined in eq.~\eqref{eq:eec}.

\section{Summary}
In this talk we have presented the NLO real-emission master integrals for the 
quark-gluon EEC in electron-positron annihilation in QCD. They are expressed
in terms of polylogarithms of two variables. 
Real-emission integrals are the most complicated pieces on the way to obtain
the complete result for EEC at NLO, hence their calculation is a crucial task.
To that end we have derived the epsilon form for the corresponding systems of
differential equation in a fully automatic manner using the implementation in
the {\sc Fuchsia} program, which is based on Lee's algorithm.
We have shown that {\sc Fuchsia} and the method are, in general, indeed
suitable to solve differential equations with multiple dimensionless variables, 
a fact that has not been mentioned in the literature so far. 

\section{Acknowledgment}
We are thankful to Vitaly Magerya for carefully reading this paper and useful comments.
The Feynman diagrams have been drawn with \texttt{Axodraw}~\cite{Ver94}.

This work has been supported by the Deutsche Forschungsgemeinschaft in
Sonderforschungs\-be\-reich 676 {\it Particles, Strings, and the Early Universe}. 


\end{document}